# Learning Inflation Narratives from Reddit: How Lightweight LLMs Reveal Forward-Looking Economic Signals

Ryuichi Saito[1,2], Sho Tsugawa[1]

[1]University of Tsukuba

[2]KU Leuven

saito.ryuichi.tkb_gw@u.tsukuba.ac.jp, s-tugawa@cs.tsukuba.ac.jp

## Abstract

Public perceptions and expectations of inflation shape household spending, wage bargaining, and policy support, making them key determinants of macroeconomic outcomes. However, current measures rely on infrequent surveys and offer limited insight into underlying narratives and sector-specific concerns. This paper presents a novel approach to measuring public perception of inflation, using lightweight large language models (LLMs) fine-tuned on domain-specific Reddit data. We created an inflation classifier trained on posts related to components of the U.S. Consumer Price Index (CPI). When applied to more than 10 years of Reddit discussions (2012–2022), this classifier produces monthly Reddit inflation scores (RIS), which we validated against actual economic indicators. Our results show that fine-tuned lightweight LLMs perform well even with smaller training datasets, and the Reddit inflation scores strongly correlate with CPI ($r = 0.91$) and closely align with the University of Michigan: Inflation Expectation (MICH). Importantly, Granger causality tests suggested that social media–based inflation scores often precede movements in both CPI and MICH, indicating their potential as predictive, forward-looking economic signals. Furthermore, change-point and lexical analyses uncovered shifts in inflation-related narratives across sectors like groceries, transportation, and housing, revealing dimensions of inflation concern that are not directly observable in aggregate price indices. By complementing traditional economic indicators with narrative-rich signals, this study demonstrates how NLP-based measures can facilitate earlier detection of inflationary pressures and policy responses.

## Introduction

Inflation impacts household finances, undermines purchasing power, and shapes consumer confidence. Empirical work shows that not only realized inflation but also households' and firms' inflation expectations affect consumption, wage setting, and support for monetary policy (Weber et al. 2022; Coibion and Gorodnichenko 2025). Evidence-based policymaking, which leverages statistical analysis to inform decision-making, is also critical for addressing complex economic challenges such as inflation. Traditionally, information on inflation and expectations has been drawn from official economic statistics and survey data. However, these sources often lag behind public perceptions and fail to capture the qualitative narratives that shape consumer behavior and expectations. Social media presents a new frontier for economic estimations using natural language processing (NLP) techniques (Guo, Hu, and Yang 2023; Leto et al. 2024; Gueta et al. 2025). Leveraging these data can provide a more immediate and nuanced understanding of public concerns, augmenting traditional metrics and enhancing the quality of economic analysis and informed policy decisions.

Although social media data offer an instantaneous and direct window into the consumer's inflation sentiment, the unstructured format presents a major challenge. First, such data are inherently fraught with noise and biases and not always representative of the specific economic situation (Einav and Levin 2014), which is a crucial difference from actual economic surveys. Second, NLP techniques, particularly large language models (LLMs) pre-trained on a vast corpus, show immense potential for inferring economic phenomena from these data, but deploying these general-purpose models incurs high computational costs. Moreover, the fine-tuning to adapt them to specific data and contexts is operationally intensive. Therefore, it is necessary to develop semantically compelling and resource-efficient methods for extracting economic signals from noisy unstructured text.

Accordingly, we focus on the U.S. economic situation and propose an approach utilizing domain-specific Reddit data and lightweight LLMs, characterized by relatively low parameters compared with foundational models such as GPT-4 and small training datasets to create a computationally efficient framework. This model provides a scalable and practical alternative. Based on the predictions, we construct a monthly Reddit Inflation Score (RIS). RIS reflects the balance between posts mentioning rising prices and those mentioning falling or stable prices, with higher scores indicating greater discussion of inflationary experiences. Additionally, we compare RIS with actual economic indicators to enhance interpretability. To this end, we investigated the following research questions.

**RQ1:** Which language models are most effective for price-related text-classification tasks, and what are their respective performance characteristics?

**RQ2:** What is the relationship between the fine-tuning dataset size and model performance, and at what point does the learning curve show diminishing returns?

**RQ3:** To what extent does the Reddit inflation score (RIS) relate to and forecast formal economic indicators?

**RQ4:** What public concerns do the RIS and the qualitative content reveal about economic phenomena?

Through this work, we provide the following contributions:

1. We design CPI-component-aligned subreddit inputs and build observational and training datasets for category-level inflation signal extraction.
2. We show that fine-tuning lightweight LLMs with domain-adaptive data results in a robust inflation index that aligns closely with, and at times precedes, official economic statistics.
3. We provide a qualitative analysis of the thematic narratives underlying public inflation sentiment, revealing specific concerns around food, cars, real estate, travel, and frugality that offer deep insights.

These contributions show that social media–based inflation indicators can deliver actionable insights. For economists, RIS transforms price narratives into a high-frequency signal to analyze expectations and category-level attention shifts. For policymakers, it serves as a proof of concept, uncovering category-specific concerns and communication gaps during price surges, complementing official statistics.

# Related Work

## Social Media Signals for Economic Measurement

Early research demonstrated that social media sentiment can reflect broad public attitudes and sometimes forecast trends in economic indicators or events (O'Connor et al. 2010; Bollen, Mao, and Pepe 2021). Subsequent studies linked text analysis to specific macroeconomic concepts such as inflation expectations and consumer confidence, including Twitter-based measures of expectations (Angelico et al. 2022) and news media–driven economic sentiment estimation during COVID-19 (Aguilar et al. 2021). Tonneau et al. (2022) employed multilingual transformer-based classification to detect personal employment-status disclosures on Twitter, addressing class imbalance via active learning. Ang and Lim (2022) predicted financial returns, volatilities, and correlations by integrating global and local multimodal data with inter-company relationships, using attention mechanisms. Guo, Hu, and Yang (2023) pointed out that these signals are prone to temporal drift as narratives change, which affects their long-term stability. Therefore, we aimed to achieve more accurate economic inference by leveraging a context-aware transformer-based language model and long-form text from Reddit.

## Forecasting Inflation using Survey Data

The econometrics literature compares neural networks and traditional models (Moshiri and Cameron 2000; Nakamura 2005), assesses machine learning for inflation forecasting in data-rich settings (Medeiros et al. 2021), and develops recurrent models for headline- and component-level Consumer Price Index (CPI) analysis (Barkan et al. 2023; Paranhos 2025). Work on explainable machine learning highlights model transparency for inflation prediction (Aras and Lisboa 2022) and includes foreign exchange reserves for emerging markets (Mirza et al. 2024). Our research uses noisy qualitative and unstructured data, incorporating it at the CPI component level in order to propose a language model specialized in inflation estimation.

## Finance and Economics with LLMs

Recent work operationalizes LLMs in economics through distinct modeling choices for task adaptation: (i) prompting/in-context learning (ICL) to extract signals or forecasts without task-specific training, (ii) agentic simulation, embedding LLMs as decision modules, and (iii) fine-tuning (including LoRA (Hu et al. 2021)) to build controllable, domain-adapted models for measurement tasks.

Faria-e-Castro and Leibovici (2024) use an LLM to generate conditional inflation forecasts from prompts and compare them with the Survey of Professional Forecasters. Boss, Longo, and Onorante (2025) prompt Llama 3 to label Reddit r/europe posts, aggregate them into euro-area indicators, and report nowcasting gains. This is analogous to our U.S.-focused setup, but we conduct a fine-grained analysis of individual CPI components and systematically examine the benefits of fine-tuning over zero-shot settings. Gueta et al. (2025) investigate LLM-based representations for financial tasks, prompting frozen models to distill social media narratives. They highlight both the potential of narrative extraction and significant challenges in quantifying economic signals for forecasting. Li et al. (2024) propose LLM agents that simulate macroeconomic activities, in which decisions rely on structured prompts and memory rather than on task-specific fine-tuning. Finally, Leto et al. (2024) quantify framing bias in U.S. news by comparing article stances with ground-truth data, demonstrating how LLMs can audit editorial choices.

Despite their practicality, prompt-only and ICL predictions remain sensitive to Reddit's noisy, unstructured narratives. We therefore evaluate supervised fine-tuning of a lightweight LLM on Reddit data to obtain stable decision boundaries. Given the lightweight model and three-way classification task, full-parameter fine-tuning is feasible, avoiding the complexity of parameter-efficient methods.

# Datasets

We filter posts using price-related keywords and classify them as inflation, deflation, or neither with a model trained on human-labeled data. The monthly RIS is computed from the shares of inflation, deflation, and neither post. This section details the construction of the dataset; model training and RIS calculation are described in the Methods section.

**Data Collection**

Reddit stands out for its communities aligned with CPI categories, which have featured price discussions for over a decade. Therefore, we collected historical posts from the Reddit archives (The-Eye n.d.) from January 2010 to December 2022. To select subreddits (discussion communities on Reddit) we referred to the CPI taxonomy in the U.S. (U.S. Bureau of Labor Statistics n.d.). The CPI is an indicator of the average change in prices for a market basket of consumer goods and services. It is recorded monthly by the Bureau of Labor Statistics. The CPI comprises some major categories, and we chose subreddits with several hundred monthly keyword hits that discussed the theme of each category. Table 1 shows all CPI components and the corresponding subreddits we adopted.

r/food is a subreddit where users post and comment on original food images. r/cars is the largest automobile-related community on the Web, covering vehicles, industry news, and personal advice. r/RealEstate covers buying homes, lending mortgages, and investing in real estate. r/travel is a community for users to talk about travel. r/Frugal is a subreddit where users exchange ideas about the lifestyle of frugality, encompassing time, money, and convenience; it covers everything from food at home to medical care services in the CPI configuration. These subreddits have a history before 2010, and contain sufficient numbers of monthly posts, even when filtered with price-related keywords. Although some subreddits are related to energy, such as r/energy and r/electricity, they were not included because their users tend not to discuss daily issues. Also, in the r/medicine, there was little discussion about price.

To collect price-related posts, the authors used search keywords such as 'price', 'cost', 'inflation', 'deflation', 'expensive', 'cheap', 'purchase', and 'sale', based on previous inflation research (Angelico et al. 2022). In r/travel, mainland U.S. identifiers were added as search keyword strings to include only U.S. destinations (Figure A1 in the Appendix).

**Observational Data**

We used observational data from January 2012 to December 2022 to analyze the perception of inflation in the U.S. (Table 2). To minimize bias in the number of posts among subreddits, if there were more than 200 posts per month for submissions, we randomly sampled 200, and if there were more than 800 comments per month, we randomly sampled 800, the ratio of which was based on the ratio between submissions and comments in each subreddit. (r/food has fewer submissions compared with the other subreddits because the posts are mainly food pictures.)

| CPI Components | Subreddits |
|---|---|
| Food | |
|   Food at home | r/food, r/Frugal |
|   Food away from home | |
| Energy | |
|   Energy commodities | r/cars, r/travel, r/Frugal |
|     Gasoline (all types) | |
|     Fuel oil | |
|   Energy services | |
|     Electricity | r/Frugal |
|     Utility (piped) gas service | |
| All items less food and energy | |
|   Commodities less food and energy commodities | |
|     New vehicles | r/cars |
|     Used cars and trucks | |
|     Apparel | r/Frugal |
|     Medical care services | |
|   Service less energy services | |
|     Shelter | r/RealEstate |
|     Transportation services | r/travel |
|     Medical care services | r/Frugal |

Table 1: Subreddits corresponding to all CPI components. Each subreddit is assumed to discuss services and goods that make up the CPI.

**Training Data**

We used data from January 2010 to December 2011 as training, validation, and test data to create an inflation classifier. These data were collected using the same search keywords as the observational data. We extracted equal numbers of submissions and comments among subreddits, so that each subreddit had an equal distribution. However, r/travel had fewer posts, and we were only able to extract 249 posts. As a result, after deduplication, a total of 1,239 posts were used for training, validation, and testing.

An author and crowdsource (Amazon Mechanical Turk) workers in the U.S. labeled the data as 0 for deflation, 2 for inflation, or 1 for neither, based on the instructions shown in Figure 1, and a majority vote for three decisions determined the final label. We restricted MTurk workers with a Master's qualification and required a ≥99% HIT approval rate and ≥5,000 approved HITs (Peer, Vosgerau, and Acquisti 2014) with no per-worker limits. For the development subset (1,239 posts; 2,478 HITs), we manually screened submissions, reposting 109 low-effort HITs and re-annotating 71 posts (including cases where labels were unchanged). These qualification filters align with common "qualified worker" strategies for MTurk reliability (Shimoni and Axelrod 2025).

|                  | r/food | r/cars  | r/RealEstate | r/travel | r/Frugal |
|------------------|--------|---------|--------------|----------|----------|
| Submissions, n   | 1,639  | 23,952  | 25,031       | 15,012   | 26,167   |
| Comments, n      | 95,729 | 105,600 | 103,380      | 54,269   | 105,600  |
| Unique users, n  | 60,569 | 49,390  | 43,072       | 33,317   | 68,628   |

Table 2: Observational data summary from subreddits between January 2012 and December 2022.

| Data types               | Deflation, n (%) | Neither, n (%) | Inflation, n (%) |
|--------------------------|------------------|----------------|------------------|
| Training data (n=782 )   | 241 (30.8)       | 300 (38.4)     | 241 (30.8)       |
| Validation data (n= 257) | 80 (31.1)        | 100 (38.9)     | 77 (30.0)        |
| Test data (n= 200)       | 62 (31.0)        | 77 (38.5)      | 61 (30.5)        |

Table 3: Training, validation, and test data used to create the inflation inference model.

---

You are a chief economist at the IMF. I would like you to infer the public perception of inflation from Reddit posts. Please classify each Reddit post into one of the following categories:

0: The post indicates deflation, such as the lower price of goods or services (e.g., "the prices are not bad"), affordable services (e.g., "this champagne is cheap and delicious"), sales information (e.g., "you can get it for only 10 dollars."), or a declining and buyer's market.

2: The post indicates or includes inflation, such as the higher price of goods or services (e.g., "it's not cheap"), the unreasonable cost of goods or services (e.g., "the food is overpriced and cold"), consumers struggling to afford necessities (e.g., "items are too expensive to buy"), shortage of goods of services, or mention about an asset bubble.

1: The post indicates neither deflation (0) nor inflation (2). This category also includes just questions to a community, social statements not personal experience, factual observations, references to originally expensive or cheap goods or services (e.g., "a gorgeous and costly dinner" or "an affordable Civic"), website promotion, authors' wishes, or illogical text.

Please choose a stronger stance when the text includes both 0 and 2 stances. If these stances are of the same degree, answer 1.

---

Figure 1: Instructions to participants for creating the training, validation, and test data, and a prompt to LLMs for inflation classification. The role as an IMF economist was specified only for the prompt. For RIS construction, we convert labels {0,1,2} to scores {−1,0,1} and compute RIS as the monthly mean.

Posts that could not be decided by a majority vote because each of the three indicated different opinions were labeled as 1 for neither. After labeling 1,239 posts, the unanimous agreement rate among all three evaluators was 67.2% (833/1,239), and the two-out-of-three agreement rate was 19.7% (244/1,239). The remaining 13.1% (162/1,239) of posts were marked as 1 for neither because they were of different values. Inter-rater reliability is reported in Appendix Table A1 (Krippendorff's α; overall α = 0.48). Labeled posts were shuffled and then filtered to extract 1,039 instances for training the language models. These were partitioned into training (75%) and validation (25%) datasets. The remaining 200 instances were reserved for testing. Table 3 shows the data details for each labeled class.

## Methods

### Pipeline Overview

Here, we present the methodological overview and corresponding research questions (Figure 2). Building on the pipeline, this section describes (i) model training and inference, (ii) evaluation and model selection, (iii) RIS construction, (iv) validation against economic indicators, and (v) change-point and lexical analyses. The classification task (Figure 1) and data construction are described in the Datasets section. The source code in this study is available at: https://github.com/RyuichiSaito1/inflation-reddit-usa

### Training of Language Models

#### Model Architecture

We built inflation classifiers with various architectures. As generative models via API, we used Gemini 2.0 Flash Lite (Google DeepMind 2025) and GPT-4.1 mini (OpenAI 2025); as open-source generative models, we used Llama 3.2 (Grattafiori et al. 2024) and Phi 2.7 (Javaheripi and Bubeck 2023); and as encoder models, we used DeBERTaV3 large (He, Gao, and Chen 2021) and RoBERTa Large (Liu et al. 2019). Gemini 2.0 Flash Lite and GPT-4.1 mini are lightweight, fast, and cost-effective alternatives to their larger, more powerful counterparts. Llama 3.2 and Phi 2.7 are open-source models with similarly sized parameters to Gemini 2.0 Flash Lite and GPT-4.1 mini. Saito and Tsugawa (2025) performed a longitudinal analysis using a tri-polar classification approach, and accuracies of over 75% were confirmed using an encoder model that was fine-tuned with domain-specific training data. Accordingly, we set encoder-based models as a baseline.

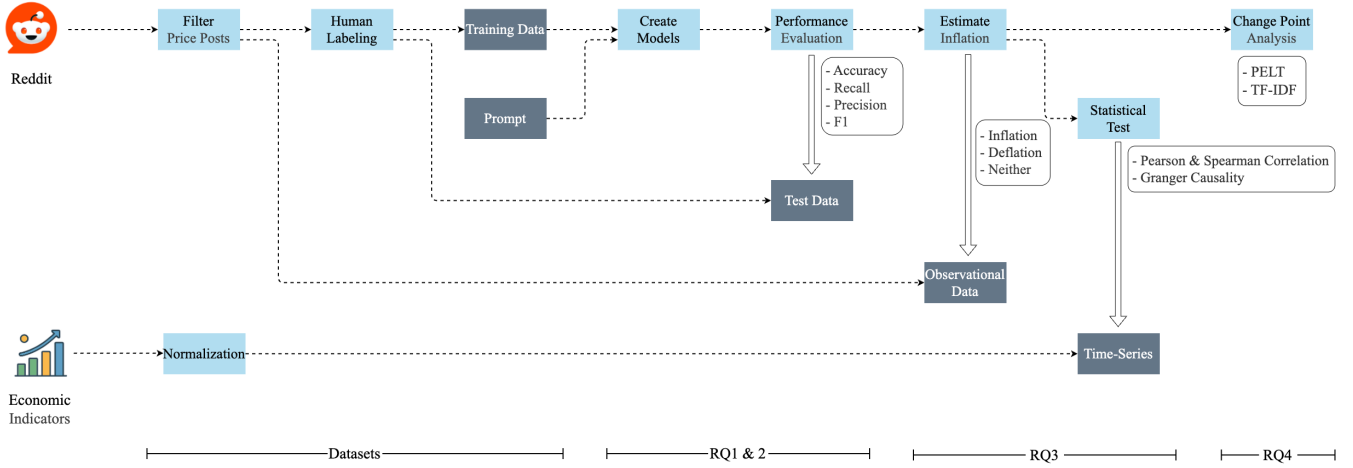

Figure 2. Research Pipeline Overview.

**Experimental Settings, Evaluation, and Model Selection**

In the performance measurement, we fine-tuned each model with 65 labeled data, which were divided into training (75%) and validation (25%) datasets, and doubled the datasets until there were 1,039 data. Furthermore, the prompt in Figure 1 was incorporated in the generative models, and the performance was examined in zero-shot (ZS) and supervised fine-tuning (SFT) settings. To evaluate the developed classification model, we used accuracy, which is the proportion of correctly predicted instances out of all the cases in the test data; recall, which is the proportion of correctly predicted instances out of all actual instances for each class (deflation, neither, and inflation); precision, which is the proportion of correctly predicted instances out of all the cases predicted as each class; and F1 score, which provides a balanced assessment by considering both recall and precision. After comparison, we selected the highest-performing model (based on the test-set metrics above) to infer inflation from the observational data.

**Reddit Inflation Score (RIS) Construction**

**Subreddit-level RIS**

The monthly RIS of a subreddit is the average inflation score of its posts within a given month. For convenience, we remapped the classification labels from {0: deflation, 1: neither, 2: inflation} to {−1: deflation, 0: neither, 1: inflation}, with the neutral class as zero. Formally, the monthly inflation score of subreddit $s$ at month $t$ was derived by

$$\overline{x}_{s,t} = \frac{1}{N_{s,t}} \sum_{i=1}^{N_{s,t}} c_{s,t,i}$$

, where $N_{s,t}$ denotes the number of posts in subreddit $s$ during month $t$, and $c_{s,t,i}$ represents the classification label of post $i$ submitted to subreddit $s$ in month $t$. A 3-month moving average is applied to smooth the monthly inflation scores. This smoothing reduces noise from unrelated posts or temporary events that do not reflect authentic economic sentiment and captures gradual inflation trends that may precede economic indicators.

**Aggregate RIS Across Subreddits**

Using the inflation scores of individual subreddits, the overall RIS for month $t$ is defined as

$$\overline{x}_t = \frac{1}{N_t} \sum_{s=1}^{S} \sum_{i=1}^{N_{s,t}} c_{s,t,i}$$

, where $N_t$ denotes the total number of posts across all subreddits in month $t$. After that, a 3-month moving average is applied to flatten the monthly inflation scores.

**Comparison Baseline**

To contextualize the LLM classifier, we compare RIS with two simple baselines constructed without supervised learning. First, a volume-based proxy: the monthly ratio of price-related posts (using our keyword filter) to total posts. Second, a rule-based sentiment baseline using VADER (Hutto and Gilbert 2014), aggregating compound scores into monthly averages. For comparability, we apply the same 3-month smoothing to these baselines and evaluate them against official inflation indicators alongside RIS.

## Time-Series Validation with Economic Indicators

### Economic Indicators and Normalization

To assess temporal relationships between the RIS and actual inflation indicators, we used the CPI and the University of Michigan: Inflation Expectation (MICH). For the CPI, we used the "All items in U.S. city average, all urban consumers, not seasonally adjusted" dataset and the original index values to ensure alignment with extracted sentiment data. MICH is a monthly consumer sentiment tool measuring anticipated inflation rates over a 1-year horizon. This forward-looking indicator captures household expectations about future price changes, derived from the university's Survey of Consumer Sentiment administered to approximately 500 consumers monthly (University of Michigan 2025).

### Correlation Analysis

We used both Pearson and Spearman correlation coefficients to assess the strength and nature of the relationship between the RIS and actual economic indicators. Pearson correlation ($r$) measures linear association between variables, with values ranging from −1 to +1. Spearman correlation ($\rho$) captures monotonic relationships, providing robustness against non-linear associations and outliers.

### Granger Causality Tests

To examine temporal precedence and predictive relationships, we used the Granger causality test (Granger 1969), which tests whether past values of one time series provide significant predictive information for another. The test yields F-statistics and corresponding p-values for both directional relationships:

$H_0$: RIS does not Granger-cause CPI or MICH
$H_0$: CPI or MICH does not Granger-cause the RIS.

## Change-Point Detection and Lexical Shift

### Change-Point Detection

To identify major discourse shifts, we apply the Pruned Exact Linear Time (PELT) algorithm (Killick, Fearnhead, and Eckley 2012) with an L2 cost to each subreddit's inflation series. We prefer this to Bayesian approaches to obtain deterministic segmentations that isolate structural breaks while remaining robust to stochastic fluctuations in social media data. We use a BIC-type penalty (Schwarz 1978), $\beta = c \cdot \log(n)\sigma^2$ (baseline $c=1.0$), where $n$ is the number of time points, and $\sigma^2$ is the variance, with a minimum segment length of m = 2. Finally, we conduct sensitivity analyses on penalty strength and segment length.

### Lexical Shift Analysis

To capture lexical change around the primary change point—the structural break consistently observed across the subreddit—we compute the difference in average bigram TF–IDF weights (Salton and Buckley 1988) between the "before" and "after" periods. We retain only inflation-labeled posts (label=2), remove URLs, and add scraping keywords to the stopword list. For each subreddit, we defined an analysis window surrounding the detected changepoint. Posts (documents) within this window were divided into "before" and "after" periods; the exact durations of these periods vary by subreddit. To ensure consistent vocabulary and comparable weights, we fitted a single TF–IDF model on the entire corpus (the union of both periods). We then computed the shift in the average weight for each bigram ($\text{Mean}_{After}$ - $\text{Mean}_{Before}$) to identify terms that characterize the transition.

## Results

### Evaluation of Classification Accuracy

#### Performance Comparison

Table 4 presents a comprehensive evaluation of six language models for inflation inference in ZS and SFT settings. Performance improved when models were adapted through SFT on domain-specific price-related datasets.

In the ZS setting, model performance differed significantly across architectures. Gemini 2.0 Flash Lite achieved the highest accuracy (0.72) and F1-score (0.73), followed by GPT-4.1 mini (both 0.66). The open-source models Llama 3.2 and Phi 2.7 had low ZS performance, both achieving accuracy and F1-scores of 0.38 and 0.33, respectively. In the ZS evaluation, the API-based model showed high performance in learning economic narratives.

In the SFT setting, most models showed improvements, with several achieving high-level performance. Gemini 2.0 Flash Lite reached an accuracy and F1-score of 0.78 after fine-tuning, improving by 0.06 and 0.05 points, respectively. GPT-4.1 mini achieved an accuracy and F1-score of 0.72 after fine-tuning, improving by 0.06 points on both metrics. Llama 3.2 demonstrated the most substantial gain, improving from 0.38 to 0.77 accuracy. Similarly, Phi 2.7 improved from 0.38 to 0.72 accuracy. Among the encoder models, RoBERTa Large and DeBERTaV3 Large also reached an accuracy and F1-score of 0.73. These results suggest that although open-source generative models exhibited substantial performance gains compared with the ZS setting, Gemini 2.0 Flash Lite maintained the highest performance. Furthermore, the encoder models, such as DeBERTaV3 Large, performed almost on par with generative models, including GPT-4.1 mini and Phi 2.7. The execution environments and hyperparameters in these models are detailed in Table A2 of the Appendix.

#### Training Efficiency

Figure 3 illustrates the relationship between training data size and model accuracy for three representative models: Gemini 2.0 Flash Lite, Llama 3.2, and DeBERTaV3. The analysis reveals different learning patterns across different model architectures. Gemini 2.0 Flash Lite exhibits data efficiency and stability. Starting from a strong ZS baseline of 0.72 accuracy, the model maintains consistently high performance across all training data sizes, with minimal variation between 0.72 and 0.78 accuracy.

| Models | Parameter | Setting | Accuracy | Precision | Recall | F1-Score |
|---|---|---|---|---|---|---|
| Gemini 2.0 Flash Lite | - | ZS | 0.72 | 0.76 | 0.73 | 0.73 |
| | | SFT | 0.78 | 0.78 | 0.79 | 0.78 |
| GPT-4.1 mini | - | ZS | 0.66 | 0.74 | 0.64 | 0.66 |
| | | SFT | 0.72 | 0.75 | 0.74 | 0.72 |
| Llama 3.2 | 3B | ZS | 0.38 | 0.37 | 0.36 | 0.33 |
| | | SFT | 0.77 | 0.77 | 0.77 | 0.77 |
| Phi 2.7 | 2.7B | ZS | 0.38 | 0.36 | 0.36 | 0.33 |
| | | SFT | 0.72 | 0.72 | 0.72 | 0.72 |
| DeBERTaV3 Large | 435M | SFT | 0.73 | 0.74 | 0.73 | 0.73 |
| RoBERTa Large | 355M | SFT | 0.73 | 0.73 | 0.73 | 0.73 |

Table 4: Performance comparison of language models for inflation inference. ZS, zero-shot; SFT, supervised fine-tuning (training data: 1,039). In the ZS setting, API-based generative models exhibited notably higher performance, and all generative models showed performance gains when fine-tuned.

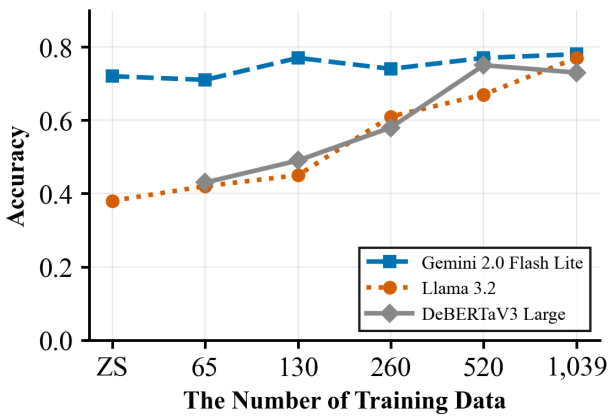

Figure 3: Training efficiency among the language models for inflation inference based on the number of training data. Gemini 2.0 Flash Lite showed consistently high performance across all training data sizes.

In contrast, Llama 3.2 has a linear learning curve that necessitates substantial domain adaptation. Beginning at 0.38 ZS accuracy, the model shows gradual improvement to 0.77 accuracy with the full 1,039 training dataset. This pattern suggests that while Llama 3.2 requires more extensive fine-tuning, it ultimately achieves competitive performance levels. DeBERTaV3 Large exhibits a similar learning pattern to Llama 3.2, starting at 0.42 with the 65 training set, with gradual improvement throughout the training progression, reaching 0.73 accuracy with the complete dataset.

Based on the comprehensive evaluation results, we selected Gemini 2.0 Flash Lite as our primary model for inflation inference tasks due to its consistently highest performance. Table A3 in the Appendix shows examples of posts and inflation-score classifications.

### Correlation Test with Economic Indicators
**Correlation with CPI**
Figure 4 presents the relationship between aggregated RIS and CPI from March 2012 to December 2022. The analysis reveals a remarkably strong positive correlation between RIS and official inflation indicators. Both time series exhibit synchronized upward trends.

The correlation tests demonstrate alignment between the measures. The Pearson correlation coefficient of $r = 0.91$ ($p < 0.001$) shows a strong linear relationship, while the Spearman correlation of $\rho = 0.89$ ($p < 0.001$) confirms a monotonic association between price-related Reddit discourse and actual price changes. The statistical significance of both measures ($p < 0.001$) suggests that the alignment is unlikely to be due to chance.

The temporal dynamics reveal several critical periods. From 2012 to 2016, both measures maintained relatively stable levels with modest upward trends. A notable acceleration period occurred from 2020 onwards. RIS and CPI increased sharply, culminating in peaks during 2021–2022. This pattern corresponds to the well-documented inflationary period following the COVID-19 pandemic (Hannon, Dube, and Xie 2021; Casselman 2024), suggesting that Reddit communities discussed inflation concerns contemporaneously.

**Correlation with MICH**
Figure 5 shows the correspondence between RIS and MICH. The correlation analysis yields a positive relationship, though weaker than the CPI comparison. The Pearson correlation of $r = 0.75$ ($p < 0.001$) indicates substantial linear association, while the Spearman correlation of $\rho = 0.20$ ($p = 0.02$) suggests a potentially non-linear relationship between the measures. The divergence between these coefficients indicates that while the variables exhibit strong linear correlation, their rank-order relationship is more variable.

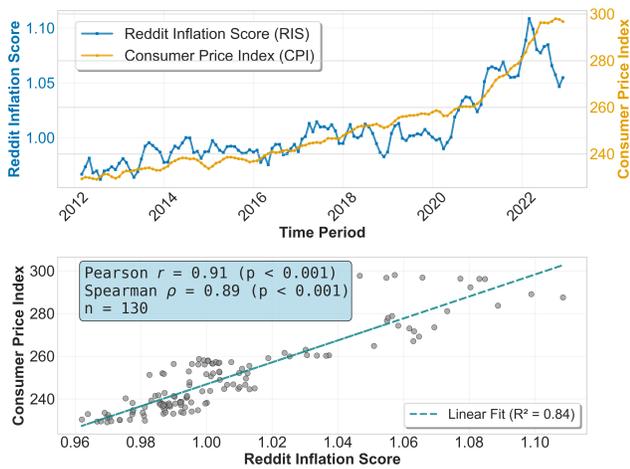

Figure 4: Reddit inflation score (RIS) vs consumer price index (CPI).

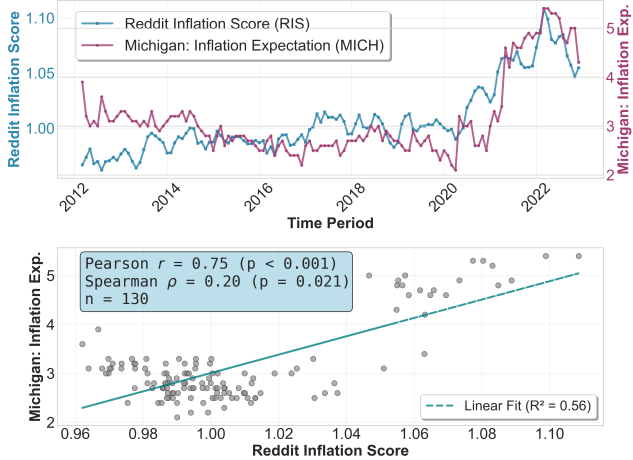

Figure 5: Reddit inflation score (RIS) vs Michigan: Inflation Expectation (MICH).

The temporal patterns reveal interesting movement between forward-looking survey expectations and social media discourse. During 2012–2015, MICH remained relatively elevated (3.0%–4.0%) while RIS showed more moderate levels. This divergence suggests that formal survey respondents expressed higher inflation concerns compared with Reddit users during this period. A remarkable convergence occurred during 2020–2022, where both measures exhibited synchronized increases. This convergence during the recent inflationary episode suggests that extreme economic conditions amplify the alignment between survey-based expectations and social media sentiment.

### Comparison with Baseline

RIS is more strongly correlated with economic indicators than two simple heuristics (volume share and VADER; see Appendix Figures A2–A3). The volume-share baseline is negatively associated with CPI (Pearson $r = -0.38$, Spearman $\rho = -0.48$; both $p < 0.001$), whereas VADER sentiment is only moderately correlated (Pearson $r = 0.53$, Spearman $\rho = 0.43$; both $p < 0.001$). These results support the claim that RIS captures inflation-relevant discourse beyond volume or sentiment polarity alone.

### Baseline Comparison: CPI vs. MICH

Figure A4 in the Appendix provides context by examining the relationship between the two official economic indicators: CPI and MICH. This baseline comparison helps interpret the relative performance of RIS. The traditional indicators demonstrate a strong positive correlation with a Pearson coefficient of $r = 0.71$ ($p < 0.001$), identical to MICH. However, the Spearman correlation of $\rho = 0.13$ ($p = 0.13$) falls below statistical significance, indicating weak rank-order correspondence between survey expectations and realized inflation.

### Causality Test with Economic Indicators

Table 5 shows bidirectional causality tests between RIS and both CPI and MICH, examining lags of 1, 2, and 3 months to capture temporal changes in economic relationships.

### Inflation Scores as Predictors of Economic Indicators

The test reveals that RIS contains significant predictive information for both traditional economic measures. For the relationship between RIS and CPI, the results demonstrate predictive causality across all examined lag structures. The strongest predictive relationship occurs with 1-month lag (F = 14.64, $p < 0.001$), showing that Reddit discussions provide significant predictive power for subsequent CPI movements. This relationship remains statistically significant with 2-month lag (F = 6.33, $p = 0.002$) and marginally so with 3-month lag (F = 3.72, $p = 0.014$).

Similarly, RIS exhibits strong predictive capacity for MICH. The 1-month lag relationship yields the highest F-statistic (F = 14.10, $p < 0.001$), suggesting that social media sentiment strongly predicts formal survey-based expectations. This predictive power persists at 2-month (F = 4.81, $p = 0.001$) and 3-month (F = 3.06, $p = 0.03$) horizons, indicating the sustained forecasting value of Reddit discussions for consumer sentiment formation.

### Economic Indicators as Predictors of Inflation Scores

In contrast, traditional economic indicators show limited ability to forecast social media inflation sentiment. Neither CPI nor MICH exhibits statistically significant Granger causality toward RIS across any examined lag. For CPI predicting RIS, all F-statistics remain low and non-significant: 1-month lag (F = 0.07, $p = 0.790$), 2-month lag (F = 1.44,

| Direction | lag | F Statistic | P value | Direction | lag | F Statistic | P value |
|---|---|---|---|---|---|---|---|
| Reddit Inflation Score (RIS) → Consumer Price Index (CPI) | 1 | 14.64 | .000 *** | Consumer Price Index (CPI) → Reddit Inflation Score (RIS) | 1 | 0.07 | .790 |
| | 2 | 6.33 | .002 ** | | 2 | 1.44 | .240 |
| | 3 | 3.72 | .014 * | | 3 | 1.06 | .369 |
| Reddit Inflation Score (RIS) → Michigan: Inflation Exp. (MICH) | 1 | 14.10 | .000 *** | Michigan: Inflation Exp. (MICH) → Reddit Inflation Score (RIS) | 1 | 1.21 | .272 |
| | 2 | 4.81 | .001 ** | | 2 | 1.34 | .266 |
| | 3 | 3.06 | .03 * | | 3 | 1.90 | .133 |

Table 5: Granger Causality test between Reddit Inflation Score (RIS) and Michigan Inflation Expectations (MICH) from March 2012 to December 2022. *, **, and *** of $P$ value indicate statistical significance at the 5%, 1%, and 0.1% levels, respectively. The arrow (→) indicates that the preceding variable is hypothesized to Granger-cause the subsequent variable

$p = 0.240$), and 3-month lag (F = 1.06, $p = 0.369$). This suggests that realized inflation rates do not systematically predict subsequent changes in social media discussion sentiment.

MICH similarly did not demonstrate predictive power for RIS, with consistently low F-statistics and non-significant $p$-values across all lag periods: 1-month (F = 1.21, $p = 0.272$), 2-month (F = 1.34, $p = 0.266$), and 3-month (F = 1.90, $p = 0.133$). Thus, formal survey-based expectations do not influence subsequent social media inflation discourse.

## Thematic Shift

### Change Point of Subreddit Communities

Figure A5 in the Appendix presents the time series of inflation scores and their change point for five subreddits, revealing distinct temporal patterns in inflation narratives. The PELT algorithm identified multiple structural breaks in each community's inflation score time series, with vertical lines marking detected change points and pink lines highlighting periods of major transitions. Sensitivity analyses varying the penalty strength and the minimum segment length show that the main changepoints are robust and do not materially affect our interpretation (Appendix Table A4).

The analysis demonstrates that different communities experienced inflation discourse shifts at varying time points, reflecting sector-specific economic concerns. r/RealEstate exhibited the highest baseline inflation scores ($\mu = 1.13$) with four major change points and r/RealEstate and r/travel showed the most volatile pattern ($\sigma = 0.05$). Notably, all communities showed pronounced increases in inflation discourse during 2020–2022, corresponding to the pandemic-era economic disruption period (Center on Budget and Policy Priorities 2022).

### Lexical Shift Analysis

Table A5 in the Appendix summarizes bigram phrases that shift most strongly across the primary changepoint, contrasting "before" and "after" terms. In r/food, the earlier period is characterized by preference- and quality-oriented phrases such as "kobe beef," "dry aged," and "looks delicious." Post-shift, the vocabulary becomes explicitly price- and affordability-oriented, featuring terms such as "beef prices," "just overpriced," and everyday items like "hot dog(s)", "chicken wings," and "pizza place," consistent with heightened price salience.

In r/cars, the pre-shift period reflects broader consumer discussion ("german cars," "daily driver," "good deal"), whereas the post-shift period foregrounds market constraints, including "used car," "chip shortage," and "gas prices," aligning with pandemic-era supply disruptions (Federal Reserve Bank of St. Louis 2025a). In r/RealEstate, the earlier period emphasizes transaction terms ("closing costs," "cash flow," "property management"), whereas the later period shifts toward macroeconomic conditions, referencing "home prices," "rising rate," and "mortgage rates," signaling a move toward rate-driven narratives.

In r/travel, the earlier period features generic planning ("make sure," "round trip"), while the later period highlights pandemic and cost constraints ("covid test," "rental prices"), reflecting mobility disruptions and sharp rental hikes (Naughton 2020). Finally, r/Frugal shifts from general budgeting ("credit card," "way cheaper") toward inflation and substitution channels ("dollar tree," "gas prices"), indicating coping strategies amid rising prices. "Buy house" also signals attention to major purchases amid soaring housing and mortgage costs (Federal Reserve Bank of St. Louis 2025b), paralleling r/RealEstate.

## Discussion

This study shows that fine-tuning lightweight LLMs on domain-specific data effectively and efficiently measures public perception of inflation. We discuss the results below in relation to our research questions and prior literature.

### Model Performance and Efficiency (RQ1 & RQ2)

The results from our model evaluation address RQ1 and RQ2, confirming the viability of lightweight LLMs for this task. We found that for API-based models such as Gemini 2.0 Flash Lite and GPT-4.1 mini, a carefully designed ZS prompt can achieve high baseline performance, indicating that these models have strong pre-trained abilities to understand economic narratives without extensive fine-tuning.

However, when given a thoughtfully labeled training dataset, even smaller, open-source generative models, such as Llama 3.2, and encoder-based models, such as DeBERTaV3 Large, showed a linear and relatively diminished return with 1,039 datasets. This suggests that API-based models are more data-efficient, while open-source generative and encoder models can reach competitive performance when supplied with a carefully labeled training dataset. This finding is important for researchers and practitioners who may lack access to large datasets. Nonetheless, our results indicate that API-based models are better suited for learning the nuances of economic narratives in classification tasks. The ability to attain high accuracy with a relatively small, manually labeled dataset (1,039 instances) demonstrates a practical and scalable approach for domain-specific NLP tasks in the social sciences. Furthermore, while Dong et al. (2023) have analyzed conversations on Reddit during the pandemic using a lexicon-based method, our study also shows the effectiveness of transformer-based language models for understanding long-form text, such as that found on Reddit.

**Relationship between Reddit Inflation Score and Economic Measurements (RQ3)**

The statistical tests on correlation and causality provide evidence for a strong link between our social media-derived inflation score and official economic indicators, answering RQ3. The remarkably high Pearson correlation with the CPI ($r = 0.91$) not only suggests a strong linear relationship but also validates our model's ability to capture real-world economic phenomena from unstructured text. This finding builds on previous work that used social media sentiment to track inflation (Angelico et al. 2022).

The Granger causality test further strengthens our claims, showing that RIS significantly Granger-causes both CPI and MICH across various lags. This result is critical for two reasons. First, it suggests that changes in public discourse on social media can precede changes in official economic measures. This is valuable for policymaking and forecasting because social media data offer real-time information that traditional indicators cannot capture. Second, the absence of a reverse causal relationship (CPI or MICH Granger-causing RIS) indicates that public discussion not only reflects a reactive echo of news releases but also provides an independent, forward-looking signal of consumer sentiment. This finding addresses a key limitation of text-driven approaches often noted in the literature, where qualitative data are not representative and distorted (Einav and Levin 2014).

**Subreddit-Specific Thematic Shifts (RQ4)**

Change-point detection and thematic analysis offer a more granular understanding of the waveforms of the inflation score. The PELT algorithm identified structural breaks in subreddit-specific inflation scores, revealing that different communities responded to economic changes at different times and in different ways. Using TF-IDF to extract bigram words, we were able to pinpoint the linguistic shifts that corresponded with major economic events, including the supply chain crisis in the car market and safety concerns about public transportation during the pandemic. This supports statistical results to analyze unstructured data.

**Real-World Implications**

Behavioral economics typically measures expectation formation and salience via surveys and experiments (DellaVigna 2009; Chetty, Looney, and Kroft 2009). RIS provides a complementary, high-frequency signal from large-scale text, continuously tracking "in the wild" price experiences, unlike episodic surveys. Our RQ3 analysis shows RIS predicts economic indicators, suggesting aggregated narratives serve as early warnings of perceived inflation. By capturing spontaneous attention, RIS may detect early expectation formation before surveys (e.g., MICH). Finally, RIS's granularity identifies which sectors (e.g., food vs. cars) drive expectations, even within the same price environment. For policy institutions, it serves as a proof of concept for monitoring category-specific concerns and identifying communication gaps during price surges. Although built on English-language Reddit, the pipeline is platform-independent and adaptable to more representative conversational data sources.

**Limitations**

This study has several limitations. The data are limited to Reddit posts, and while we chose price-related subreddits, the findings may not be generalizable to national-level demographics. Research in computational social science has highlighted the platform-specific and population bias inherent in online platforms (Ruths and Pfeffer 2014). The user base of platforms such as Reddit does not reflect the general population, skewing toward demographic groups, including younger, male, and technologically savvy individuals. Moreover, reliance on explicit price-related keywords may underrepresent implicit inflation perceptions lacking price terms. Further, our framework leverages Reddit's CPI-aligned structure; adapting it to unstructured platforms (e.g., X and Facebook) requires tailored collection strategies. The annotation agreement was also moderate (Krippendorff's α < 0.667; Table A1), indicating ambiguity in classifying posts and introducing label noise that may have affected SFT more than ZS, because only SFT was directly trained on the annotated labels. Finally, the Granger causality test suggests a predictive relationship but does not imply structural causality, and further research is needed to elucidate the underlying mechanisms of why public discourse precedes economic indicators. Future work should explore a multi-modal approach that integrates both qualitative social media data and quantitative economic variables to build more robust models.

## Conclusion

This study presents a novel method for extracting economic signals from social media, thereby enhancing our understanding of macroeconomic phenomena. We showed that by fine-tuning a lightweight LLM on a carefully crafted, domain-specific dataset, a reliable measure of public inflation sentiment can be developed. Our work makes a key contribution by connecting unstructured, qualitative online data with established, quantitative economic indicators, thereby supporting the use of computational methods in economic analysis.

The primary impact of our research is the establishment of a validated, forward-looking indicator. The strong correlation ($r = 0.91$) with CPI validated our model's ability to capture real-world price pressures. More significantly, our Granger Causality analysis revealed that the social media-derived inflation score precedes shifts in both CPI and MICH. This finding is pivotal because it elevates social media data from a mere reactive echo of economic news to an independent, predictive signal. This capability directly addresses the challenge of immediacy in traditional economics, where key indicators are often published with a lag.

By providing this interpretable and predictive framework, our research advances computational social science. It offers a replicable blueprint for transforming noisy, unstructured conversations into structured insights, enhancing the legitimacy of using online data for economic and policy-relevant analysis. Future research should aim to generalize this approach by incorporating data from diverse platforms and international contexts, and by exploring multi-modal models. Ultimately, this work substantiates the power of modern NLP to distill actionable macroeconomic intelligence.

## Acknowledgements

This work was supported by JST SPRING Grant No. JPMJSP2124 and JSPS KAKENHI Grant No. JP25K03105.

# Paper Checklist

1. For most authors...
   (a) Would answering this research question advance science without violating social contracts, such as violating privacy norms, perpetuating unfair profiling, exacerbating the socio-economic divide, or implying disrespect to societies or cultures? Yes
   (b) Do your main claims in the abstract and introduction accurately reflect the paper's contributions and scope? Yes
   (c) Do you clarify how the proposed methodological approach is appropriate for the claims made? Yes
   (d) Do you clarify what are possible artifacts in the data used, given population-specific distributions? Yes
   (e) Did you describe the limitations of your work? Yes
   (f) Did you discuss any potential negative societal impacts of your work? N/A due to economic phenomenon analysis
   (g) Did you discuss any potential misuse of your work? N/A due to economic phenomenon analysis
   (h) Did you describe steps taken to prevent or mitigate potential negative outcomes of the research, such as data and model documentation, data anonymization, responsible release, access control, and the reproducibility of findings? Yes, see Ethical Statement
   (i) Have you read the ethics review guidelines and ensured that your paper conforms to them? Yes
2. Additionally, if your study involves hypotheses testing...
   (a) Did you clearly state the assumptions underlying all theoretical results? Yes
   (b) Have you provided justifications for all theoretical results? Yes
   (c) Did you discuss competing hypotheses or theories that might challenge or complement your theoretical results? Yes
   (d) Have you considered alternative mechanisms or explanations that might account for the same outcomes observed in your study? Yes
   (e) Did you address potential biases or limitations in your theoretical framework? Yes
   (f) Have you related your theoretical results to the existing literature in social science? Yes
   (g) Did you discuss the implications of your theoretical results for policy, practice, or further research in the social science domain? Yes
3. Additionally, if you are including theoretical proofs...
   (a) Did you state the full set of assumptions of all theoretical results? N/A
   (b) Did you include complete proofs of all theoretical results? N/A
4. Additionally, if you ran machine learning experiments...
   (a) Did you include the code, data, and instructions needed to reproduce the main experimental results (either in the supplemental material or as a URL)? Yes, after completing the peer review process, we will publish our code and data in the GitHub repository.
   (b) Did you specify all the training details (e.g., data splits, hyperparameters, how they were chosen)? Yes
   (c) Did you report error bars (e.g., with respect to the random seed after running experiments multiple times)? N/A
   (d) Did you include the total amount of compute and the type of resources used (e.g., type of GPUs, internal cluster, or cloud provider)? Yes
   (e) Do you justify how the proposed evaluation is sufficient and appropriate to the claims made? Yes
   (f) Do you discuss what is "the cost" of misclassification and fault (in)tolerance? N/A
5. Additionally, if you are using existing assets (e.g., code, data, models) or curating/releasing new assets, without compromising anonymity...
   (a) If your work uses existing assets, did you cite the creators? Yes
   (b) Did you mention the license of the assets? No, because the datasets used in this paper have no explicit license statements.
   (c) Did you include any new assets in the supplemental material or as a URL? N/A
   (d) Did you discuss whether and how consent was obtained from people whose data you're using/curating? N/A
   (j) Did you discuss whether the data you are using/curating contains personally identifiable information or offensive content? Yes, see Ethical Statement
   (e) If you are curating or releasing new datasets, did you discuss how you intend to make your datasets FAIR (see FORCE11 (2020))? N/A
   (f) If you are curating or releasing new datasets, did you create a Datasheet for the Dataset (see Gebru et al. (2021))? N/A
6. Additionally, if you used crowdsourcing or conducted research with human subjects, without compromising anonymity...
   (a) Did you include the full text of instructions given to participants and screenshots? Yes

(b) Did you describe any potential participant risks, with mentions of Institutional Review Board (IRB) approvals?
We used crowdsourced annotators for text labeling. Under our institution's human-subjects research policy, this study was deemed low risk and did not require formal IRB review; therefore, no IRB approval number was issued.

(c) Did you include the estimated hourly wage paid to participants and the total amount spent on participant compensation? Yes, see Ethical Statement

(d) Did you discuss how data is stored, shared, and deidentified? N/A

## Ethical Statement

### Data Privacy and Anonymization

All Reddit posts used in this study were publicly accessible and obtained from archived datasets. Since public posts may still contain personally identifiable information, we do not distribute raw post content, author handles, or post-level labels. For reproducibility, we provide only limited metadata necessary to retrieve the original records, and our analyses emphasize overall patterns rather than individual users.

### Human Subject Protection

Amazon MTurk workers participated voluntarily in the annotation task. Workers with master qualifications living in the U.S. were selected to ensure high-quality annotations while providing fair compensation for skilled work. The estimated hourly wage was $15.12 USD, and the total amount spent on compensation was $298.8 USD. No sensitive personal information was requested from annotators.

### IRB / Ethics Review Status

Under our institution's human-subjects research policy, this study was deemed low risk and did not require formal IRB review; therefore, no IRB approval number was issued.

# Appendix

**Datasets**

Figure A1 lists the comprehensive set of search terms associated with geographic entities to filter the dataset in r/travel. Table A1 presents the statistical assessment of inter-annotator consistency across categories in the annotation tasks.

---

**Synonyms of the U.S.**

' america ', 'united states', ' usa ', 'the us', 'u.s.', 'stateside', 'across the states'

**U.S. States**

'alabama', 'alaska', 'arizona', 'arkansas', 'california', 'colorado', 'connecticut', 'delaware', 'florida', 'georgia', 'hawaii', 'idaho', 'illinois', 'indiana', 'iowa', 'kansas', 'kentucky', 'louisiana', 'maine', 'maryland', 'massachusetts', 'michigan', 'minnesota', 'mississippi', 'missouri', 'montana', 'nebraska', 'nevada', 'new hampshire', 'new jersey', 'new mexico', 'new york', 'north carolina', 'north dakota', 'ohio', 'oklahoma', 'oregon', 'pennsylvania', 'rhode island', 'south carolina', 'south dakota', 'tennessee', 'texas', 'utah', 'vermont', 'virginia', 'washington', ' DC ', 'west virginia', 'wisconsin', 'wyoming'

**Major U.S. Cities by Population (Top 50)**

'new york', ' ny ', ' nyc ', 'los angeles', 'chicago', 'houston', 'phoenix', 'philadelphia', 'san antonio', 'san diego', 'dallas', 'san jose', 'austin', 'jacksonville', 'fort worth', 'columbus', 'indianapolis', 'charlotte', 'san francisco', 'seattle', 'nashville', 'denver', 'oklahoma city', 'el paso', 'boston', 'portland', 'las vegas', 'vegas', 'detroit', 'memphis', 'louisville', 'baltimore', 'milwaukee', 'albuquerque', 'tucson', 'fresno', 'sacramento', 'kansas city', ' mesa ', 'atlanta', 'omaha', 'colorado springs', 'raleigh', 'long beach', 'virginia beach', 'miami', 'oakland', 'minneapolis', 'tulsa', 'bakersfield', 'wichita', 'arlington', 'aurora', 'tampa', 'new orleans', 'cleveland', 'honolulu', 'anaheim', 'lexington', 'stockton', 'corpus christi', 'henderson', 'riverside', 'newark', 'st. paul', 'santa ana', 'cincinnati', 'irvine', 'orlando', 'pittsburgh', 'st. louis', 'greensboro', 'jersey city', 'anchorage', 'lincoln', 'plano', 'durham', 'buffalo', 'chandler', 'chula vista', 'toledo', 'madison', 'gilbert', ' reno ', 'fort wayne', 'north las vegas', 'st. petersburg', 'lubbock', 'irving', 'laredo', 'winston-salem', 'chesapeake', 'glendale', 'garland', 'scottsdale', 'norfolk', 'boise', 'fremont', 'spokane', 'santa clarita', 'baton rouge', 'richmond', 'hialeah'

**Major Tourist Spots**

'grand canyon', 'yellowstone', 'hollywood', 'niagara', 'disney world', 'yosemite', 'central park'

**Transportation Services**

'amtrak', 'greyhound', 'interstate'

---

Figure A1: Additional search keywords for data collection in r/travel to include only U.S. destination posts.

| Subreddit | Number | Alpha |
|---|---|---|
| r/food | 245 | 0.61 |
| r/cars | 249 | 0.44 |
| r/RealEstate | 248 | 0.46 |
| r/travel | 249 | 0.39 |
| r/Frugal | 248 | 0.48 |
| Total | 1,239 | 0.48 |

Table A1. Inter-rater reliability (Krippendorff's α) of the annotation task. α tends to be higher in r/food (with more explicit price mentions) and lower in r/travel, where prices.

## Evaluation of Classification Accuracy

Table A2 outlines the computational environments, hyperparameter settings, and training results used to develop classification models. Table A3 presents representative examples of categorized Reddit posts, along with their assigned inflation scores and definitions.

| Models | Platforms | Train Batch Size | Val Batch Size | Total Epoch | Selected Epoch | Learning Rate | Train Loss | Val Loss |
|---|---|---|---|---|---|---|---|---|
| Gemini 2.0 Flash Lite | Google Cloud Vertex AI | Not specified | Not specified | 8 | 8 | Auto* | 0.19 | 0.24 |
| GPT-4.1 mini | Open AI API | 1 | Not specified | 3 | 2 | Auto* | 1.40 | 0.59 |
| Llama 3.2 | Google Colab A100 GPU | 4 | 4 | 4 | 3 | $5 \times 10^{-5}$ | 0.61 | 0.87 |
| Phi 2.7 | Google Colab A100 GPU | 4 | 4 | 10 | 7 | $2 \times 10^{-5}$ | 1.10 | 0.64 |
| DeBERTaV3 Large | Google Colab A100 GPU | 8 | 4 | 6 | 5 | $2 \times 10^{-5}$ | 0.42 | 0.94 |
| RoBERTa Large | Google Colab L4 GPU | 8 | 4 | 5 | 6 | $2 \times 10^{-5}$ | 0.17 | 1.37 |

Table A2: Environment and hyperparameters in creating models. For Gemini 2.0 Flash Lite and GPT-4.1 mini, hyperparameters were automatically optimized by the respective API services based on dataset size and characteristics. The final models were selected from epochs where both training and validation losses indicated optimal convergence, ensuring robust generalization performance while avoiding overfitting.

*The absolute learning rates are determined by the managed services. Values indicate the multipliers applied to these baselines: 0.8 for Gemini and 2.0 for GPT.

| Posts | Inflation Score | Definition |
|---|---|---|
| I have a bag of green spirulina that I got for very cheap (~$4) and I was really excited about using it in smoothies. I tried it in one and the taste is definitely not for me, but I'm so desperate to use it up. Does anyone have ideas for how I can mask the very fishy flavor of spirulina? Recipes and ideas greatly appreciated :) | −1 | Deflation |
| Maryland here and I have bought new plants every spring for the past few years BUT this summer we planted 2 rosemary bushes in the shade of towering Jerusalem artichokes and they really boomed. I may have a shot with these... but for sure, Chicago is a different story. | 0 | Neither |
| We are still about 6 months out from our home construction being completed. We are increasingly worried that we will be priced out of the home by the time next year comes. We stretched out a little this time around so the mortgage was affordable just on the higher side of the budget. But now we're not to sure if this going to be fatal plus the additional increases still to come. Any thoughts on this scenario? Being priced out during the construction phase due to the rates. Thanks | 1 | Inflation |

Table A3: Sample of posts and inflation score classified using fine-tuned Gemini 2.0 Flash Lite. For convenience, we remapped the classification labels from {0: deflation, 1: neither, 2: inflation} to {−1: deflation, 0: neither, 1: inflation}, with the neutral class as zero. This mapped score (−1/0/1) is the value averaged to compute RIS.

## Correlation Test with Economic Indicators

Figure A2 shows a longitudinal comparison of the keyword-based volume share relative to the CPI. Figure A3 also presents a longitudinal comparison of VADER-based sentiment scores with the CPI. Additionally, Figure 4 shows the longitudinal relationship and correlation between Michigan Inflation Expectations (MICH) and the Consumer Price Index (CPI).

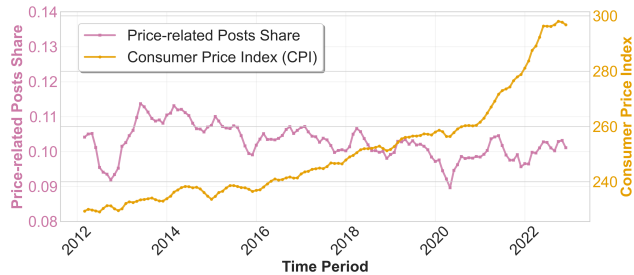

Figure A2:   Keyword-based volume-share baseline vs. consumer price index (CPI). Pearson $r$=−0.38, Spearman $\rho$=−0.48 (both $p$<0.001).

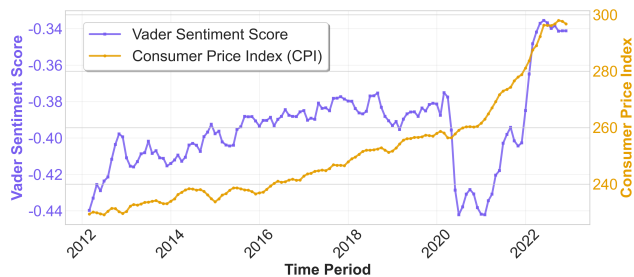

Figure A3: VADER-based sentiment baseline (monthly average compound score) vs. consumer price index (CPI). Pearson $r$=0.53, Spearman $\rho$=0.43 (both $p$<0.001).

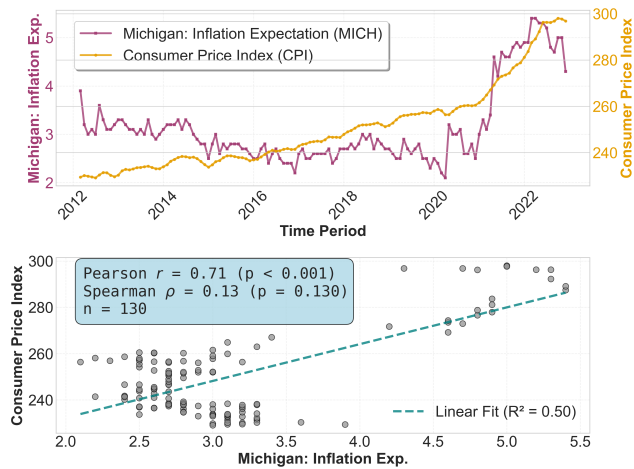

Figure A4: Reddit inflation score (RIS) vs Michigan: Inflation Expectation (MICH). Pearson $r$=0.71 ($p$<0.001), Spearman $\rho$=0.13 ($p$=0.13)

# Thematic Shift

Figure A5 shows the longitudinal evolution of inflation scores and detected structural breaks (change points) for each subreddit. Table A4 also presents the sensitivity analysis of change-point detection, listing the number and dates of identified structural breaks for each subreddit across varying penalty-parameter settings. Further, Table A5 presents the top-ranked bigrams for each subreddit, separated into periods before and after the identified structural breaks. This comparison illustrates the thematic evolution of the discourse by highlighting the terms that became dominant in each respective timeframe.

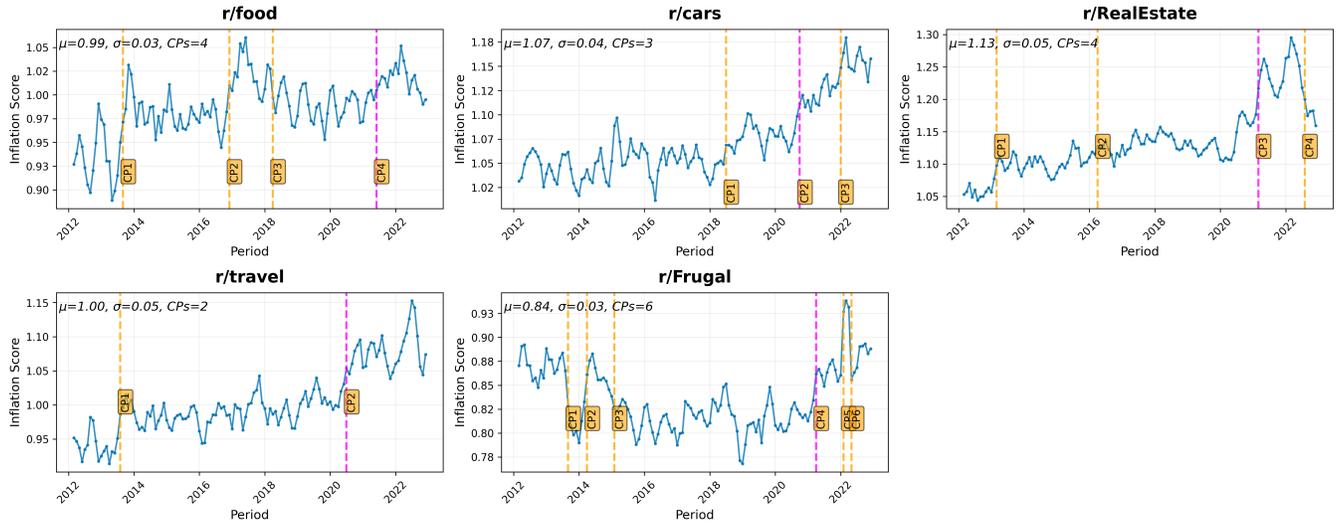

Figure A5: Inflation score and change point of each subreddit between March 2012 and December 2022. CP, change point. Pink lines highlight periods of major transitions. Inflation scores in all subreddits showed pronounced increases during 2020–2022, corresponding to the period of economic disruption during the pandemic.

| Subreddit | c (penalty) | Min. seg. (m) | No. of CPs | CP dates (YYYY-MM) |
|---|---|---|---|---|
| r/food | 0.5 | 2 | 6 | 2012-12; 2013-03; 2013-08; 2016-12; 2018-04; 2021-06 |
|  | 1.0 | 2 | 4 | 2013-09; 2016-12; 2018-04; 2021-06 |
|  | 2.0 | 2 | 2 | 2013-09; 2016-12 |
| r/cars | 0.5 | 2 | 3 | 2018-07; 2020-10; 2022-01 |
|  | 1.0 | 2 | 3 | 2018-07; 2020-10; 2022-01 |
|  | 2.0 | 2 | 2 | 2018-07; 2020-10 |
| r/RealEstate | 0.5 | 2 | 5 | 2013-02; 2015-06; 2020-07; 2021-03; 2022-08 |
|  | 1.0 | 2 | 4 | 2013-03; 2016-04; 2021-03; 2022-08 |
|  | 2.0 | 2 | 2 | 2015-06; 2021-03 |
| r/travel | 0.5 | 2 | 5 | 2013-08; 2019-02; 2020-07; 2022-05; 2022-10 |
|  | 1.0 | 2 | 2 | 2013-08; 2020-07 |
|  | 2.0 | 2 | 2 | 2013-08; 2020-07 |
| r/Frugal | 0.5 | 2 | 6 | 2013-09; 2014-04; 2015-02; 2021-04; 2022-02; 2022-05 |
|  | 1.0 | 2 | 6 | 2013-09; 2014-04; 2015-02; 2021-04; 2022-02; 2022-05 |
|  | 2.0 | 2 | 2 | 2013-09; 2021-04 |

Table A4: Sensitivity of PELT changepoints to the penalty factor and minimum segment length. For each subreddit, the table reports the number of detected changepoints (No. of CPs) and their dates (YYYY–MM) for each penalty factor $c$ (shown for $m = 2$ months; varying $m \in \{1,3\}$ yields only minor differences that do not affect our substantive conclusions). The factor c scales the BIC-type penalty $\beta = c \log(n)\sigma^2$ used in PELT.

| r/food | | r/cars | | r/RealEstate | |
|---|---|---|---|---|---|
| April 2018 to May 2021 (before) | June 2021 to December 2022 (after) | July 2018 to September 2020 (before) | October 2020 to December 2021 (after) | May 2016 to February 2021 (before) | March 2021 to July 2022 (after) |
| kobe beef | years ago | german cars | used car | closing costs | home prices |
| bit pricey | deep dish | daily driver | years ago | make sure | house prices |
| pretty good | beef prices | really nice | car prices | dont know | higher rates |
| looks delicious | hot dogs | want know | car market | dont want | rising rates |
| dry aged | holy shit | good deal | chip shortage | sellers market | housing prices |
| dont think | really good | car miles | don know | im sure | housing market |
| almond flour | hot dog | let know | used market | cash flow | prices going |
| make sure | dim sum | just buy | gas prices | bay area | low rates |
| ive wanting | pizza place | id say | current market | property management | appraisal gap |
| grocery stores | chicken wings | work vehicle | sports cars | lot money | raising rates |
| ice cream | just overpriced | ford explorer | fun cars | youre going | mortgage rates |
| vanilla extract | prices going | gas mileage | low mileage | market value | months ago |
| people pay | local butcher | car going | make sense | new roof | prices higher |
| place called | cast iron | fuel prices | vast majority | buyers agent | rates going |
| quite pricey | lot people | lose money | hot hatches | make money | rates rising |

| r/travel | | r/Frugal | |
|---|---|---|---|
| August 2013 to June 2020 (before) | July 2020 to December 2022 (after) | February 2015 to March 2021 (before) | April 2021 to January 2022 (after) |
| dont know | years ago | dont want | years ago |
| im sure | rental car | dont know | prices going |
| make sure | covid test | make sure | dollar tree |
| youre going | pcr test | ive seen | don know |
| im looking | don want | im sure | facebook marketplace |
| im going | don know | sales tax | used car |
| youre looking | tipping culture | youre paying | little pricey |
| hong kong | don think | credit card | ve seen |
| round trip | car rentals | im just | supply chain |
| let know | rental cars | long time | don need |
| new zealand | ll need | im looking | air fryer |
| ive heard | just don | im going | month groceries |
| spend days | key west | youre going | car prices |
| ive looked | rental prices | thats just | raise prices |
| im really | car prices | way cheaper | gas prices |

Table A5: Top bigrams with the largest shifts in TF–IDF weights around the change point. We compute the difference in mean weights (Mean$_{after}$ − Mean$_{before}$). The "After" column lists terms with the largest positive differences (emerging terms), while the "Before" column lists terms with the largest negative differences (declining terms). To ensure robustness, terms appearing in fewer than 5 documents or in more than 95% of the documents were excluded.

19